\documentclass[a4paper,12pt]{article}
\usepackage{cite}
\newcommand{\be}{\begin{equation}}
\newcommand{\ee}{\end{equation}}
\newcommand{\bea}{\begin{eqnarray}}
\newcommand{\nn}{\nonumber}
\newcommand{\eea}{\end{eqnarray}}

\begin{document}

\begin{titlepage}
\begin{flushright}
UB-ECM-PF-03/10
\end{flushright}
\begin{centering}
\vspace{.3in}
{\Large{\bf Cardy-Verlinde Formula and  Ach\'ucarro-Ortiz Black
Hole}}
\\

\vspace{.5in} {\bf Mohammad R. Setare$^{1}$ and Elias C.
Vagenas$^{2}$ }\\
\vspace{.3in} $^{1}$\, Department of Physics, Sharif University of
Technology, Tehran, Iran\\and\\Physics Department (Farmanieh)\\
Institute for Studies in Theoretical Physics \& Mathematics
(IPM)\\ 19395-5531 Tehran, Iran\\and\\Department of Science,
Physics group, Kordestan University, Sanandeg, Iran
\\rezakord@yahoo.com\\
\vspace{0.4in}

$^{2}$\, Departament d'Estructura i Constituents de la Mat\`{e}ria\\
and\\ CER for Astrophysics, Particle Physics and Cosmology\\
Universitat de Barcelona\\
Av. Diagonal 647, E-08028 Barcelona\\
Spain\\
evagenas@ecm.ub.es\\
\end{centering}

\vspace{0.6in}
\begin{abstract}

In this paper it is shown that the entropy of the black hole
horizon in the Ach\'ucarro-Ortiz  spacetime, which is the most
general two-dimensional black hole derived from the
three-dimensional rotating BTZ black hole, can be described by the
Cardy-Verlinde formula. The latter is supposed to be an entropy
formula of conformal field theory in any dimension.

\end{abstract}
\end{titlepage}

\newpage

\baselineskip=18pt
\section*{Introduction}
Holography is believed to be one of the fundamental principles of
the true quantum theory of gravity \cite{{HOL},{RAP}}. An
explicitly calculable example of holography is the much--studied
anti-de Sitter (AdS)/Conformal Field Theory (CFT) correspondence.
\par Lately, it has been proposed the duality of quantum gravity, in
a de Sitter (dS) space, to a certain
 Euclidean  CFT, living on a spacelike boundary of the
dS space~\cite{Strom} (see also earlier works \cite{mu2}-\hspace{-0.9ex} \cite{Bala}). This duality is defined in
analogy to the AdS$_{d}$/CFT$_{d-1}$ correspondence. Following this idea, several works on dS space have been
carried out
\cite{Mazu}-\hspace{-0.9ex} \cite{set3}.\\
The Cardy-Verlinde formula proposed by Verlinde \cite{Verl}, relates the entropy of a  certain CFT with its total
energy and its Casimir energy in arbitrary dimensions. Using the AdS$_{d}$/CFT$_{d-1}$ and dS$_{d}$/CFT$_{d-1}$
correspondences, this formula has been shown to hold exactly for the cases of Schwarzschild-dS,
Reissner-Nordstr\"om-dS, Kerr-dS, dS Kerr Newman-dS and topological dS black holes. In this paper, by using the
Cardy-Verlinde formula, we have re-obtained the entropy of the Ach\'ucarro-Ortiz black hole which is a
two-dimensional black hole derived from the three-dimensional rotating BTZ black hole.
\par In 1992
Ba\~nados, Teitelboim and Zanelli (BTZ) \cite{banados1,banados2}
showed that $(2+1)$-dimensional gravity has a black hole
 solution. This black hole is described by two (gravitational) parameters,
the mass $M$ and the angular momentum (spin) $J$. It is locally AdS and thus it differs from Schwarzschild and
Kerr solutions since it is asymptotically anti-de-Sitter instead of flat spacetime. Additionally, it has no
curvature singularity at the origin.
 AdS black holes, are members of this two-parametric family
of BTZ black holes and they are very  interesting in the framework
of string theory and black hole physics
\cite{strominger1,strominger2}.
\par For systems that
admit 2D CFTs as duals, the Cardy formula \cite{cardy} can be
applied directly. This formula gives the entropy of a CFT in terms
of the central charge $c$ and the eigenvalue of the Virasoro
operator $l_{0}$. However, it should be pointed out that this
evaluation is possible as soon as one has explicitly shown (e.g
using the AdS$_{d}$/CFT$_{d-1}$ correspondence) that the system
under consideration is in correspondence with a 2D CFT
\cite{CM99,andy}.
\par\noindent Even in the most favorable  case presented above,
the use of the Cardy formula for the computation of the entropy of
the gravitational system is far from trivial, since the central
charge $c$ and the eigenvalue $l_{0}$ of the Virasoro operator
have to be expressed in terms of the gravitational parameters
($M$, $J$)\cite{CCCM}.
\par The two-dimensional (2D) limit of the Cardy-Verlinde
proposal is interesting for various reasons.  The main motivation to study 2D black holes and 2D gravity is to use
them as the useful laboratory for more complicated 4D cousins \cite{{fab1},{fab2},{fab3}}. On the same time, quite
much is known about (basically quantum) Cardy-Verlinde formula in four dimensions. For instance, the quantum
origin for 4D entropy in relation with Cardy-Verlinde  formula is explicitly discussed in number of works
\cite{{noj1},{noj2},{br}}. Using the AdS$_{d}$/CFT$_{d-1}$ correspondence, it is known that there are 2D
gravitational systems that admit 2D CFTs as duals \cite{CM99,cadcav}. As mentioned before, one can make direct use
of the original Cardy formula \cite{cardy} to compute the entropy \cite{CM99,cadcav,med1}. A comparison of this
result with the corresponding one derived from a 2D generalization of the Cardy-Verlinde formula could be very
useful in particular for understanding of the puzzling features of the AdS$_{d}$/CFT$_{d-1}$ correspondence in two
dimensions \cite{strominger}. \par\noindent Also of great interest in extending the Cardy-Verlinde formula to
$d=2$ is the clarification of the meaning of the holographic principle for 2D spacetimes. The boundaries of
spacelike regions of 2D spacetimes are points therefore the notion  of holographic bound is far from trivial.
Furthermore, a generalization of the work of Verlinde to two spacetime dimensions presents several difficulties,
essentially for dimensional reasons. Since the black hole horizons are isolated points, one cannot establish an
area law \cite{carta1}. Additionally, the spatial coordinate is not a ``radial'' coordinate due to a scale
symmetry and hence one cannot impose a natural normalization on it \cite{carta2}.
\section{Ach\'ucarro-Ortiz  Black Hole}
 The black hole
solutions of Ba\~nados, Teitelboim and Zanelli
\cite{banados1,banados2} in $(2+1)$ spacetime dimensions are
derived from a three dimensional theory of gravity \be S=\int
dx^{3} \sqrt{-g}\,({}^{{\small(3)}} R+2\Lambda) \ee with a
negative cosmological constant ($\Lambda=\frac{1}{l^2}>0$).
\par\noindent
The corresponding line element is \be ds^2 =-\left(-M+
\frac{r^2}{l^2} +\frac{J^2}{4 r^2} \right)dt^2
+\frac{dr^2}{\left(-M+ \displaystyle{\frac{r^2}{l^2} +\frac{J^2}{4
r^2}} \right)} +r^2\left(d\theta -\frac{J}{2r^2}dt\right)^2
\label{metric}\ee There are many  ways to reduce the three
dimensional BTZ black hole solutions to the two dimensional
charged and uncharged dilatonic black holes \cite{ortiz,lowe}. The
Kaluza-Klein reduction of the $(2+1)$-dimensional metric
(\ref{metric}) yields a two-dimensional line element:
 \be ds^2 =-g(r)dt^2 +g(r)^{-1}dr^2
\label{metric1}\ee where \be g(r)=\left(-M+\frac{r^2}{l^2}
+\frac{J^2}{4 r^2}\right)\label{metric2}
 \ee
with $M$ the Arnowitt-Deser-Misner (ADM) mass, $J$ the angular momentum (spin)
 of the BTZ black hole and $-\infty<t<+\infty$, $0\leq r<+\infty$, $0\leq \theta <2\pi$.
\par \noindent
The outer and inner horizons, i.e. $r_{+}$ (henceforth simply black hole horizon) and $r_{-}$ respectively,
concerning the positive mass black hole spectrum with spin ($J\neq 0$) of the line element (\ref{metric}) are
given as  \be r^{2}_{\pm}=\frac{l^2}{2}\left(M\pm\sqrt{M^2 - \displaystyle{\frac{J^2}{l^2}} }\right)
\label{horizon1} \ee and therefore, in terms of the inner and outer horizons, the black hole mass and the angular
momentum are given, respectively, by \be M=\frac{r^{2}_{+}}{l^{2}}+\frac{J^{2}}{4r^{2}_{+}}\label{mass}\ee and \be
J=\frac{2\, r_{+}r_{-}}{l}\label{ang}\ee with the corresponding angular velocity to be \be\Omega=\frac{J}{2
r^{2}}\label{angvel}\hspace{1ex}.\ee
\par\noindent The Hawking temperature $T_H$ of the black hole
horizon is \cite{kumar1} \bea T_H &=&\frac{1}{2\pi
r_{+}}\sqrt{\left(\displaystyle{\frac{
r_{+}^2}{l^2}+\frac{J^2}{4r_{+}^2}}\right)^2-\displaystyle{\frac{J^2}{l^2}}}\nn\\
&=&\frac{1}{2\pi r_{+}}\left(\displaystyle{\frac{
r_{+}^2}{l^2}-\frac{J^2}{4r_{+}^2}}\right)\label{temp1}
\hspace{1ex}.\eea \par\noindent The area $\mathcal{A}_H$ of the
black hole horizon is \bea \mathcal{A}_H &=& 2\pi l
\left(\frac{M+\sqrt{M^2 -\frac{J^2}{l^2} }}{2}\right)^{1/2}\label{area1}\\
&=& 2\pi r_{+} \label{area2}\eea and thus the entropy of the two-dimensional Ach\'ucarro-Ortiz black hole, if we
employ the well-known Bekenstein-Hawking area formula ($S_{BH}$) for the entropy
\cite{bekenstein1,bekenstein2,hawking3}, is given by  \be S_{bh}=\frac{1}{4\hbar G} \mathcal{A}_H
=S_{BH}\hspace{1ex}. \ee Using the BTZ units where $8\hbar G =1 $, the entropy of the two-dimensional
Ach\'ucarro-Ortiz black hole  takes the form \be S_{bh}=4 \pi r_{+} \label{entr1}\hspace{1ex}.\ee
\section{Cardy-Verlinde Formula}
In a recent paper, Verlinde \cite{Verl} propound a generalization of the Cardy formula which holds for the ($1+1$)
dimensional Conformal Field Theory (CFT), to $(n+1)$-dimensional spacetime described by the metric \be
ds^{2}=-dt^{2}+R^{2}d\Omega_{n}\ee where $R$ is the radius of a $n$-dimensional sphere.
\par\noindent The generalized
Cardy formula (hereafter named Cardy-Verlinde formula)  is given by \be S_{CFT}=\frac{2\pi
R}{\sqrt{ab}}\sqrt{E_{C}\left(2E-E_{C}\right)} \label{cvf}\ee where $E$ is the total energy and $E_{C}$ is the
Casimir energy. The definition of the Casimir energy is derived by the violation of the Euler relation as \be
E_{C}\equiv n\left(E+pV-TS-\Phi Q\right)\label{casimir1}\ee where the pressure of the CFT is defined as $p=E/nV$.
The total energy may be written as the sum of two terms \be E(S, V)=E_{E}(S, V)+\frac{1}{2}E_{C}(S,
V)\label{ext}\ee where $E_{E}$ is the purely extensive part of the total energy $E$. The Casimir energy $E_{C}$ as
well as the purely extensive part of energy $E_{E}$ expressed in terms of the radius $R$ and the entropy $S$ are
written as \bea
E_{C}&=&\frac{b}{2\pi R}S^{1-\frac{1}{n}}\label{cftcas1}\\
E_{E}&=&\frac{a}{4\pi R}S^{1+\frac{1}{n}}\label{exten1}\hspace{1ex}.\eea After the work of Witten on
AdS$_{d}$/CFT$_{d-1}$ correspondence \cite{witten},  Savonije and Verlinde proved that the Cardy-Verlinde formula
(\ref{cvf}) can be derived using the thermodynamics of  AdS-Schwarzschild black holes in arbitrary dimension
\cite{sav}.
\section{Entropy of Ach\'ucarro-Ortiz black hole  in Cardy-Verlinde Formula}
We would like to derive the entropy of the two-dimensional
Ach\'ucarro-Ortiz black hole (\ref{entr1}) from the Cardy-Verlinde
formula (\ref{cvf}).
 First, we evaluate the Casimir energy $E_{C}$ using (\ref{casimir1}).
 It is easily seen from (\ref{temp1}) and (\ref{entr1}) that
 \be
 T_{H}S_{bh}=2\left(\displaystyle{\frac{
r_{+}^2}{l^2}-\frac{J^2}{4r_{+}^2}}\right)\label{ts}
 \ee
while from (\ref{ang}) and (\ref{angvel}) we have \be
\Omega_{+}J=\frac{J^2}{2r^{2}_{+}}\label{jo}\hspace{1ex}.\ee Since
the two-dimensional Ach\'ucarro-Ortiz black hole
 is asymptotically anti-de-Sitter, the total energy is $E=M$ and
thus the Casimir energy, substituting (\ref{mass}), (\ref{ts}) and (\ref{jo}) in (\ref{casimir1}), is given as \be
E_{C}=\frac{J^2}{2r_{+}^2}\label{casimir2}\ee where in our analysis the charge $Q$ is the angular momentum $J$ of
the two-dimensional Ach\'ucarro-Ortiz black hole, the corresponding electric potential $\Phi$ is the angular
velocity $\Omega$ and $n=1$. Making use of expression (\ref{cftcas1}), Casimir energy $E_{C}$ can also be written
as \be E_{C}=\frac{b}{2\pi R}\label{cftcas2}\hspace{1ex}.\ee Additionally, it is obvious that the quantity
$2E-E_{C}$ is given, by substituting (\ref{ts}) and (\ref{jo}) in (\ref{casimir1}), as \be 2E-E_{C}=2\frac{
r_{+}^2}{l^2}\label{e-ec}\hspace{1ex}.\ee The purely extensive part of the total energy $E_{E}$ by substituting
(\ref{e-ec}) in (\ref{ext}), is given as \be E_{E}= \frac{r_{+}^2}{l^2}\label{exten2}\ee whilst by substituting
(\ref{entr1}) in (\ref{exten1}), it takes the form \be E_{E}=\frac{4 \pi a
}{R}r^{2}_{+}\label{exten3}\hspace{1ex}.\ee At this point it is useful to evaluate the radius $R$. By equating the
right hand sides of  (\ref{casimir2}) and (\ref{cftcas2}), the radius is written as \be R=\frac{b r^{2}_{+}}{\pi
J^2}\label{radius1}\ee while by equating the right hand sides (\ref{exten2}) and (\ref{exten3}) it can also be
written as \be R= 4 \pi a l^2\label{radius2}\hspace{1ex}.\ee Therefore, the radius expressed in terms of the
arbitrary positive coefficients $a$ and $b$ is \be R= 2
r_{+}\left(\frac{l}{J}\right)\sqrt{ab}\label{radius3}\hspace{1ex}.\ee Finally, we substitute  expressions
(\ref{casimir2}), (\ref{e-ec}) and (\ref{radius3}) which were derived in the context of thermodynamics of the
two-dimensional Ach\'ucarro-Ortiz  black hole, in the Cardy-Verlinde formula (\ref{cvf}) which in turn was derived
in the context of CFT  \be S_{CFT}=\frac{2\pi}{\sqrt{ab}}2 r_{+}
\left(\frac{l}{J}\sqrt{ab}\right)\sqrt{\frac{J^2}{2r^{2}_{+}}\frac{2r^{2}_{+}}{l^2}}\ee and we get \be
S_{CFT}=S_{bh}\hspace{1ex}.\ee It has been proven that the entropy of the the two-dimensional Ach\'ucarro-Ortiz
black hole can be expressed in the form of Cardy-Verlinde formula.
\section{Conclusions}
Among the family of $AdS_{d}/CFT_{d-1}$ dualities, the pure gravity case $AdS_3/CFT_2$ is the best understood. In
contrast, although some progress has been made in understanding AdS/CFT correspondence in two space-time
dimensions \cite{{CM99},{strominger}},  $AdS_2/CFT_1$  remains quite enigmatic. The aim of this paper is to
further investigate the $AdS_2/CFT_1$ correspondence in terms of Cardy-Verlinde entropy formula. Naively, one
might expect that holographic dualities in a two-dimensional bulk context would be the simplest cases of all. This
may certainly be true on a calculational level; however, one finds such two-dimensional dualities to be plagued
with conceptually ambiguous features \cite{med2}. One of the remarkable outcomes of the AdS/CFT and dS/CFT
correspondence has been the generalization of Cardy's formula (Cardy-Verlinde formula) for arbitrary
dimensionality as well as for a variety of AdS and dS backgrounds. In this paper, we have shown that the entropy
of the black hole horizon of Ach\'ucarro-Ortiz spacetime can also be rewritten in the form of Cardy-Verlinde
formula.
\section{Acknowledgments}
The work of E.C.V. has been supported by the European Research and Training Network ``EUROGRID-Discrete Random
Geometries: from Solid State Physics to Quantum Gravity" (HPRN-CT-1999-00161).

\end{document}